\documentclass[twocolumn,showpacs]{revtex4}

\usepackage{graphicx}%
\usepackage{dcolumn}
\usepackage{amsmath}

\begin{document}

\title{Self-aligned fabrication process for silicon quantum computer devices}

\author{T.M. Buehler$^{1}$, R.P. McKinnon$^{1}$, N.E. Lumpkin$^{1}$, R.
Brenner$^{1}$,\\
D.J. Reilly$^{1}$, L.D. Macks$^{1}$, A.R. Hamilton$^{1}$, A.S.
Dzurak$^{2}$ and R.G. Clark$^{1}$}

\address{Centre for Quantum Computer Technology}
\address{1) School of Physics, University of New South Wales, Sydney 2052,
AUSTRALIA}
\address{2) School of Electrical Engineering, University of New
South Wales, Sydney 2052, AUSTRALIA}
\date{\today}

\begin{abstract}
We describe a fabrication process for devices with few quantum
bits (qubits), which are suitable for proof-of-principle
demonstrations of silicon-based quantum computation. The devices
follow the Kane proposal to use the nuclear spins of $^{31}$P
donors in $^{28}$Si as qubits, controlled by metal surface gates
and measured using single electron transistors (SETs). The
accurate registration of $^{31}$P donors to control gates and
read-out SETs is achieved through the use of a self-aligned
process which incorporates electron beam patterning, ion
implantation and triple-angle shadow-mask metal evaporation.
\end{abstract}


\maketitle

\begin{figure}[b]
\centering
\includegraphics[width=7.5cm]{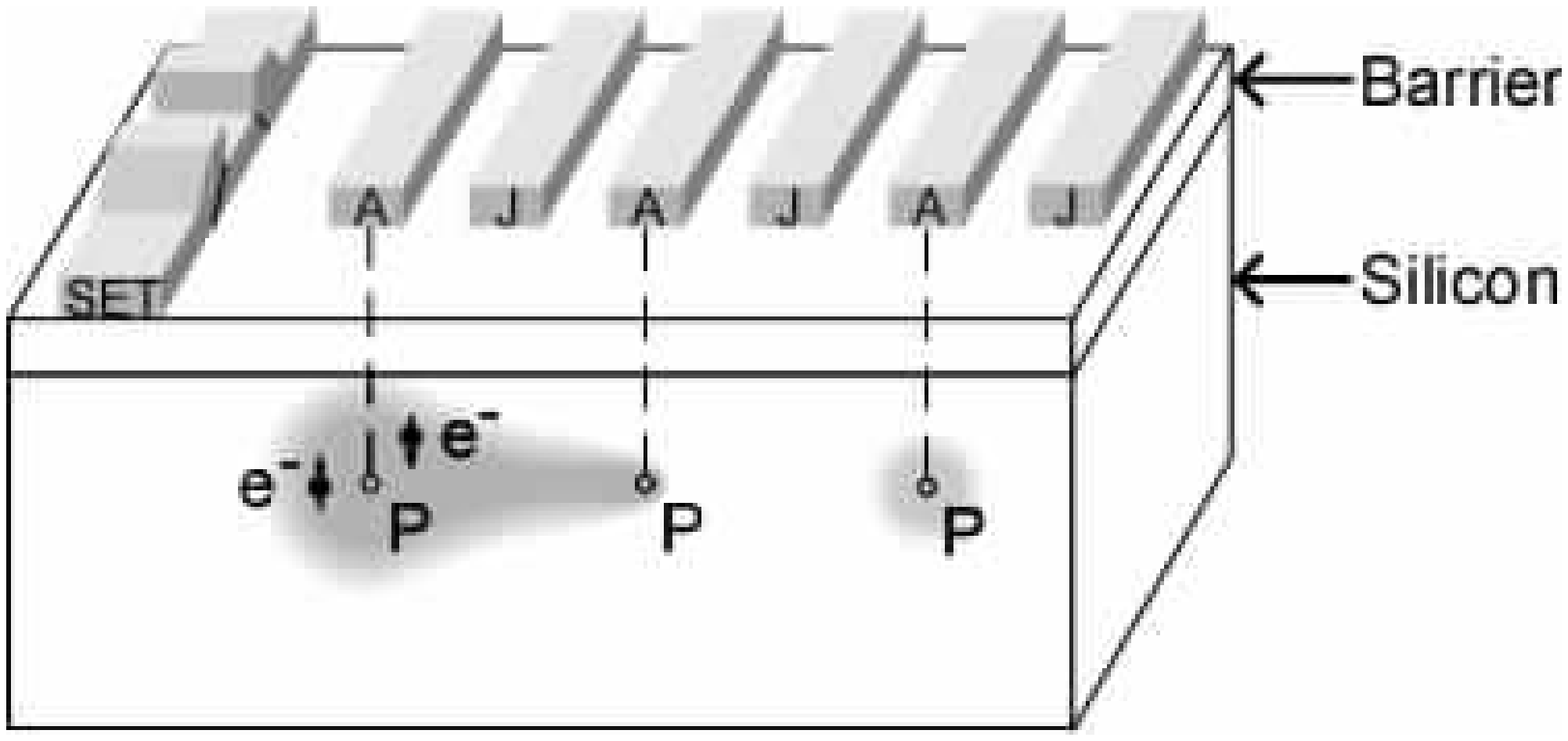}
\caption{Figure 1. SSQC architecture in which $^{31}$P atoms form
an array of quantum bits in $^{28}$Si. The SET is used to read out
the spin configuration of the two left-most donors.}
\end{figure}

Large-scale quantum computers \cite{a1} promise a massive increase
in the speed of important computational tasks such as prime
factorisation \cite{a2} and database searching \cite{a3}. One of
the most promising schemes, which is capable of being scaled up to
processors with many quantum bits (qubits), is the silicon-based
solid-state quantum computer (SSQC), first proposed by Kane
\cite{a4}. Such devices require the positioning of single dopant
atoms to better than 10 nm: a highly challenging task. However the
paradigm has a major advantage over competing proposals in that
the fabrication strategy is compatible with industry-standard
silicon metal-oxide-semiconductor (MOS) processing technology.

A silicon-based SSQC is depicted schematically in Figure 1. The
nuclear spin (I) of single $^{31}$P donor atoms (I=1/2) embedded
in $^{28}$Si (I=0) constitutes the qubit. 'A'-gates above the
donors enable single-qubit operations by controlling the hyperfine
interaction between electron and nucleus. 'J'-gates between donors
enable two-qubit operations by controlling the exchange
interaction between adjacent donors \cite{a4}. During the qubit
read-out process, nuclear spin information is transferred to
electron spin states as described by Kane \cite{a4}. Under the
influence of an asymmetric gate bias, the Pauli exclusion
principle dictates that electron motion between two P donors (to
form a two-electron bound state - see Fig. 1) will only take place
if the electrons are in a spin-singlet state. Qubit read-out is
achieved using a charge-sensitive single electron transistor (SET)
to detect whether or not this charge transfer occurs, thereby
allowing determination of the two-electron spin state \cite{a4}.
The device must be operated at temperatures of 1 K, or below, in
order to minimise qubit decoherence and to ensure that all
electron spins are in their ground state.

Realization of the structure shown in Fig. 1 requires not only the
ability to form ordered arrays of $^{31}$P atoms in $^{28}$Si, but
also a capability to accurately align these atoms to their control
gates and read-out SETs. These requirements remain relevant to
variations on the original Kane design, including those utilizing
either nuclear-spin qubits \cite{a5} or electron-spin qubits
\cite{a6}. Recently, it has been demonstrated \cite{a7} that P
qubit arrays on Si can be fabricated using an atom-structuring
approach in which a scanned probe is used to pattern a monohydride
resist on a silicon surface. While this approach can be extended
to produce high-precision P donor arrays on a large scale, a
number of challenging steps, including subsequent epitaxial
overgrowth and gate registration, have yet to be demonstrated.

This paper describes an approach to SSQC fabrication which has two
key advantages over other schemes: (i) all of its steps are based
upon existing semiconductor processes; and (ii) it employs a
self-alignment technique for registering single donors to control
gates and read-out SETs. The scheme described here is designed for
few-qubit devices which will be critical for determining the
feasibility of a SSQC. It is also possible to envisage
modifications which would allow scale-up of the process for
many-qubit array fabrication. Here we demonstrate the fabrication
of an integrated array of control gates and SETs, and present a
detailed strategy for self-aligning these structures to individual
P atoms.

\begin{figure}[t]
\centering
\includegraphics[width=7.5cm]{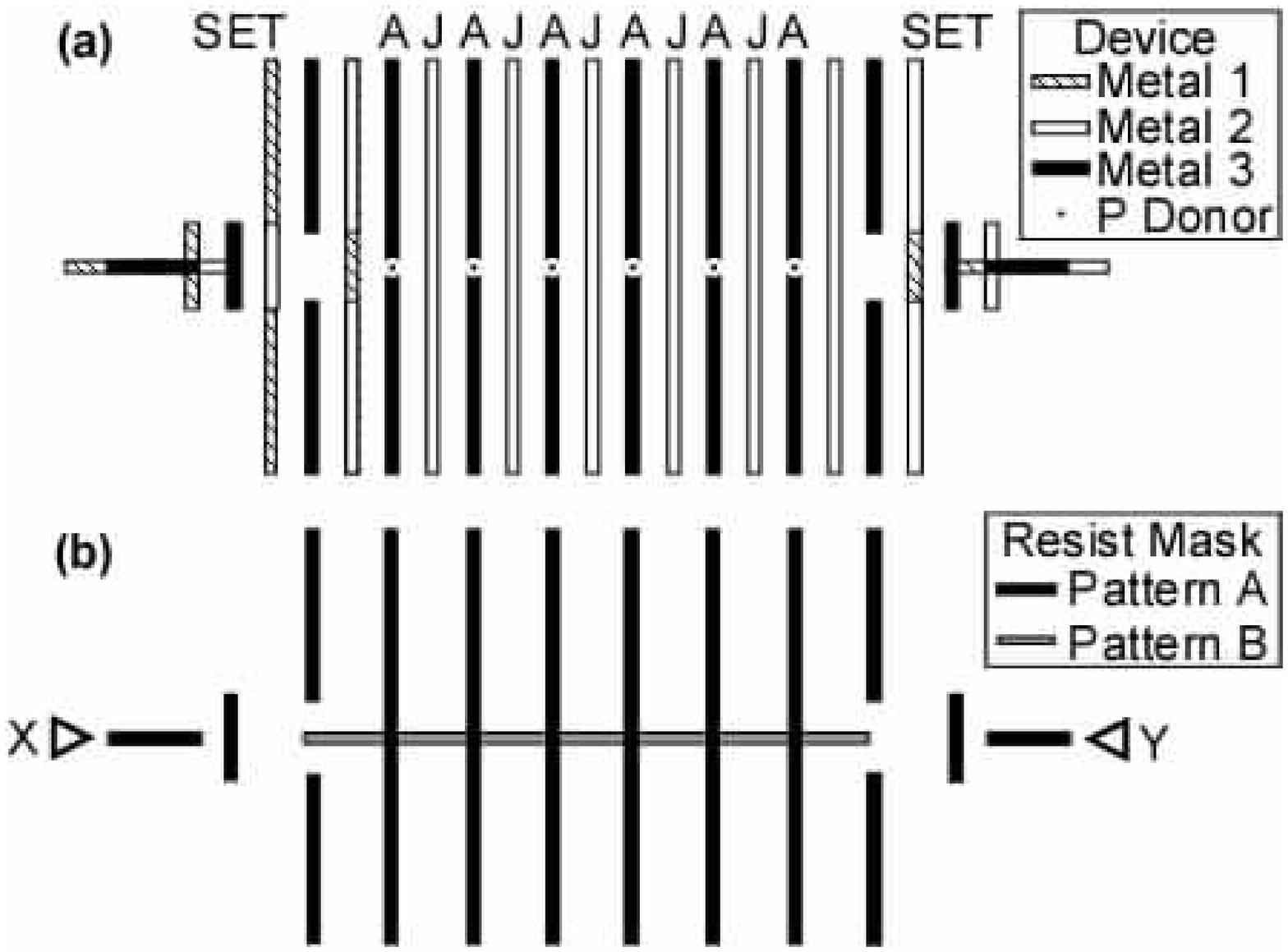}
\caption{Figure 2. (a) Schematic of a six-donor device. The dots
represent individual P atoms. (b) The two EBL patterns required to
form the structure shown in (a). The black pattern forms the gates
and SETs after triple-angle shadow evaporation, while the overlap
of the grey and black patterns defines the qubit locations.}
\end{figure}

For the purposes of this paper, we consider a six-donor,
four-qubit device as shown in Figure 2(a). We first outline the
conceptual approach for the process, then focus on a particular
implementation employing poly-methylmethacrylate (PMMA),
poly-methacrylic acid (PMAA) and polymethylglutarimide (PMGI)
resists.

Surface metallisations for the device are defined using
multi-angle shadow evaporation, which is already a
well-established technique for fabricating Al/Al$_{2}$O$_{3}$ SETs
\cite{a8}. A single high-resolution electron beam lithography
(EBL) step is used to create the black pattern (Pattern A) shown
in Fig. 2(b). Following resist development, triple-angle shadow
evaporation \cite{a9} and oxidation are performed to produce Al
and Al/Al$_{2}$O$_{3}$ structures which constitute all of the
gates and SETs shown in Fig. 2(a). Although some extraneous
features are produced by the shadow evaporation process, these
will not significantly affect the performance of the device. The
resist structure used for the shadow evaporation step (Resist A)
must be able to produce a wide cavity beneath fine features in a
surface layer \cite{a9} and would generally be comprised of two or
more resist layers of differing sensitivity, as discussed later.

A linear array of self-aligned donors can be realized by
incorporating an additional EBL step using a different resist
(Resist B). The locations of the six donors in the device may be
defined using the grey pattern (Pattern B) shown in Fig 2(b). The
points at which Patterns A and B cross define the donor sites.
Access channels for ion implantation, which extend to the
substrate at these donor sites, may be opened by either wet
development or dry etching. A critical requirement is that the two
resist structures utilize different developer solutions or
different dry etch processes.

\begin{figure}[b]
\centering
\includegraphics[width=7.5cm]{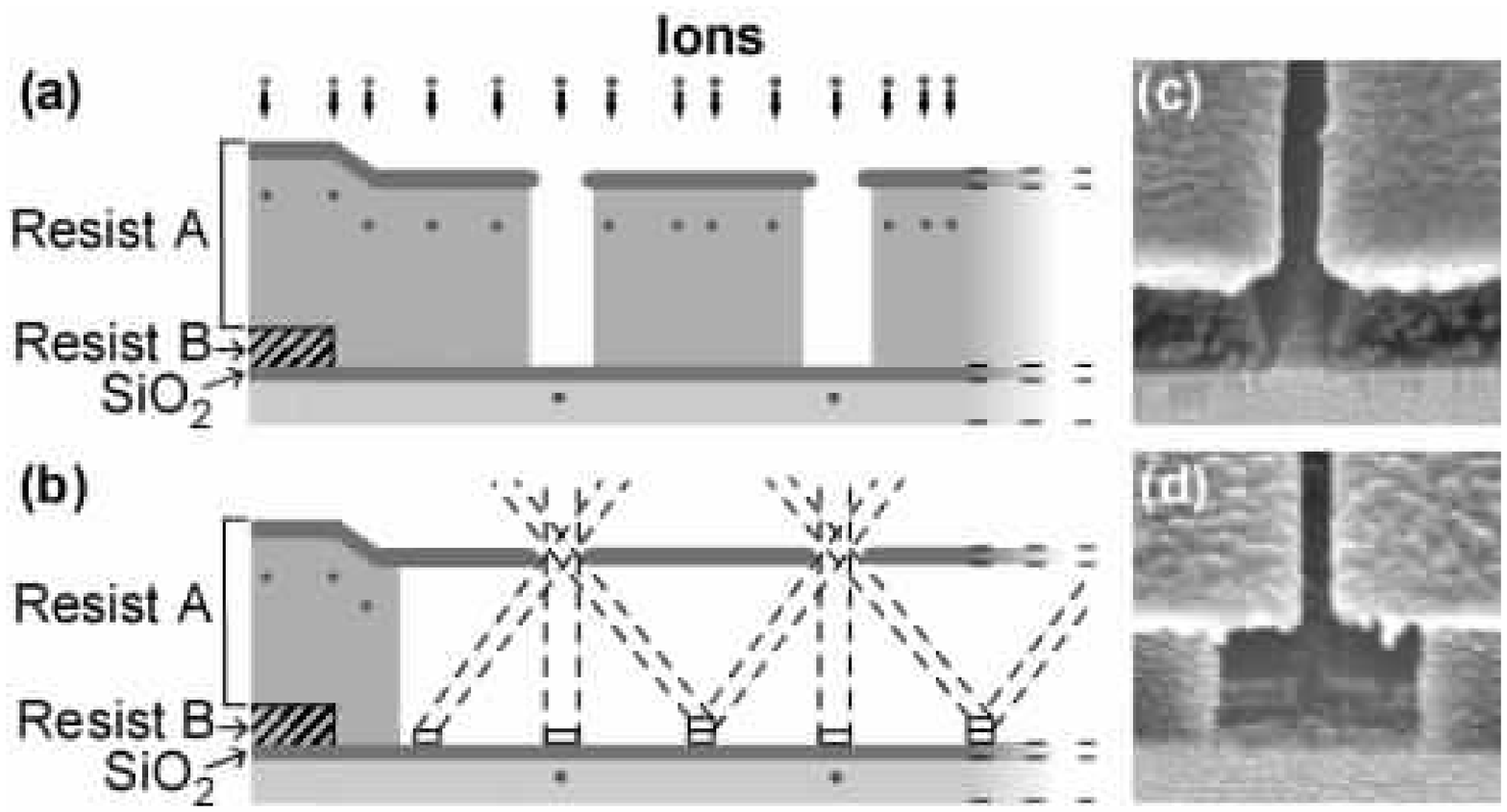}
\caption{Figure 3. (a) Resist profile for the ion implantation
step. (b) Profile for the triple-angle evaporation process.
Profiles in (a) and (b) are cross-sections along XY in Fig. 2(b).
(c) SEM image of ion implantation channel in bilayer resist of
PMMA (upper) and PMMA/PMAA (lower). (d) SEM image of multi-angle
evaporation cavity in bilayer resist.}
\end{figure}

Partial etching or development of Resist A, along with complete
etching or development of Resist B produces a profile such as that
shown in Figure 3(a) for ion implantation. Ions must be implanted
at low energy (of order keV) to ensure donors are located 5-20 nm
below the Si/SiO$_{2}$ interface and with an areal dose such that
on average one ion is implanted at each donor site. By Poissonian
statistics, the probability that a single ion lands in each of N
sites is given \cite{a6} by 0.367$^{N}$. As N increases, the
device yield would fall to low levels without additional
strategies aimed at controlling ion placement. One such strategy
involves the use of on-chip ion impact detectors which
electrically register the arrival of each ion in the device,
during the implant process \cite{a10}. This approach ensures that
exactly N donors will be implanted in an intended N-donor device.
Recent results using this technique demonstrate the detection of
single $^{31}$P+ ions of energy 15 keV entering an i-Si substrate
\cite{a10}. This ion energy leads to an average implant depth of
20 nm, suitable for a SSQC device. In order to scale-up this
process for the construction of many-qubit processors it will be
necessary to localise the ion beam to a spot size smaller than the
spacing between donor sites and move the beam between each site
until a single ion implant event is registered. Such localisation
could be achieved using a focused ion beam (FIB) or a moveable
mask with a nanomachined aperture \cite{a11}.

Although ion implantation is a convenient technique for
introducing donors into Si, there are two aspects of the process
which could impair the reliable operation of a SSQC, namely
straggle of P donors from their desired locations and damage to
the Si lattice. The effect of straggle has been estimated using
SRIM software (Stopping and Range of Ions in Matter) and for
typical donor depths of 10 nm we find a lateral straggle of order
5 nm \cite{a12}. These values suggest that it will be necessary to
individually tune the gate voltages that control each donor qubit,
to take into account its specific location. Lateral straggle can
be improved somewhat by channelling the implant along a crystal
axis, however, the amorphous SiO$_{2}$ barrier layer will serve to
randomise the incident angle and reduce channelling.

There is limited information on damage induced by $single$
implanted donors in Si, although molecular dynamics calculations
\cite{a13} indicate that considerable damage may be automatically
removed by self-annealing in the first picosecond after impact.
Residual damage can be removed by annealing after the
metallisation of the gates and removal of the resist layers. In
the worst case, it may be necessary to anneal at temperatures as
high as 900°C in order to remove impact damage and activate the P
donor electrons. Given that SET tunnel barriers are likely to be
degraded above 400°C during long anneal steps, it may be necessary
to use a rapid thermal anneal (RTA) process for these devices.
Alternatively, SETs could be fabricated in a separate EBL step
after annealing \cite{a10}. Although the alignment of SETs to
donors would be compromised in this case, read-out is still
expected to be successful due to the demonstrated high sensitivity
of SETs. An advantage of post-fabrication of the SETs is that
refractory metals could be used for the gates. Some refractory
metals, such as tungsten, have high melting points and also
exhibit low diffusivity through SiO$_{2}$ barrier layers.

Following ion implantation, complete development of Resist A
produces a cavity suitable for multi-angle shadow evaporation (see
Fig. 3b). Any remaining Resist B on the floor of the cavity is
also removed at this point. The triple-angle shadow evaporation
process depicted in Fig. 3(b) produces self-aligned gate and SET
structures with A-gates located directly above implanted donors,
thus realizing the complete device shown in Fig. 2(a).

We now describe a particular example of the fabrication strategy
outlined above which employs only wet chemical processing.
Firstly, PMGI (Resist B) is prepared on a Si/SiO$_{2}$ substrate.
Pattern B is exposed using EBL and developed with an aqueous
developer. Next, a bilayer (Resist A) of PMMA and PMMA/PMAA
copolymer is prepared on top of the PMGI and Pattern A is exposed.
Partial development of the bilayer with an organic solvent
(MIBK/IPA) is then carried out to create narrow trenches down to
the PMGI. This developer solution does not affect the PMGI. Fig.
3(c) shows the profile of a partially developed bilayer cavity.
Where the two EBL patterns overlap, channels extending to the
substrate surface are created, through which ion implantation may
be performed.

As a trial of this process, we have exposed and developed a series
of lines in bilayer resist, on top of and perpendicular to a
series of lines exposed and developed in PMGI (Fig. 4a). The
resulting structure was etched using an HF solution and then the
resist layers removed. Atomic force microscope (AFM) imaging
(Figs. 4b and 4c) confirmed that etch pits in the Si/SiO$_{2}$
substrate were formed only where the two sets of lines overlapped.

\begin{figure}[h]
\centering
\includegraphics[width=7.5cm]{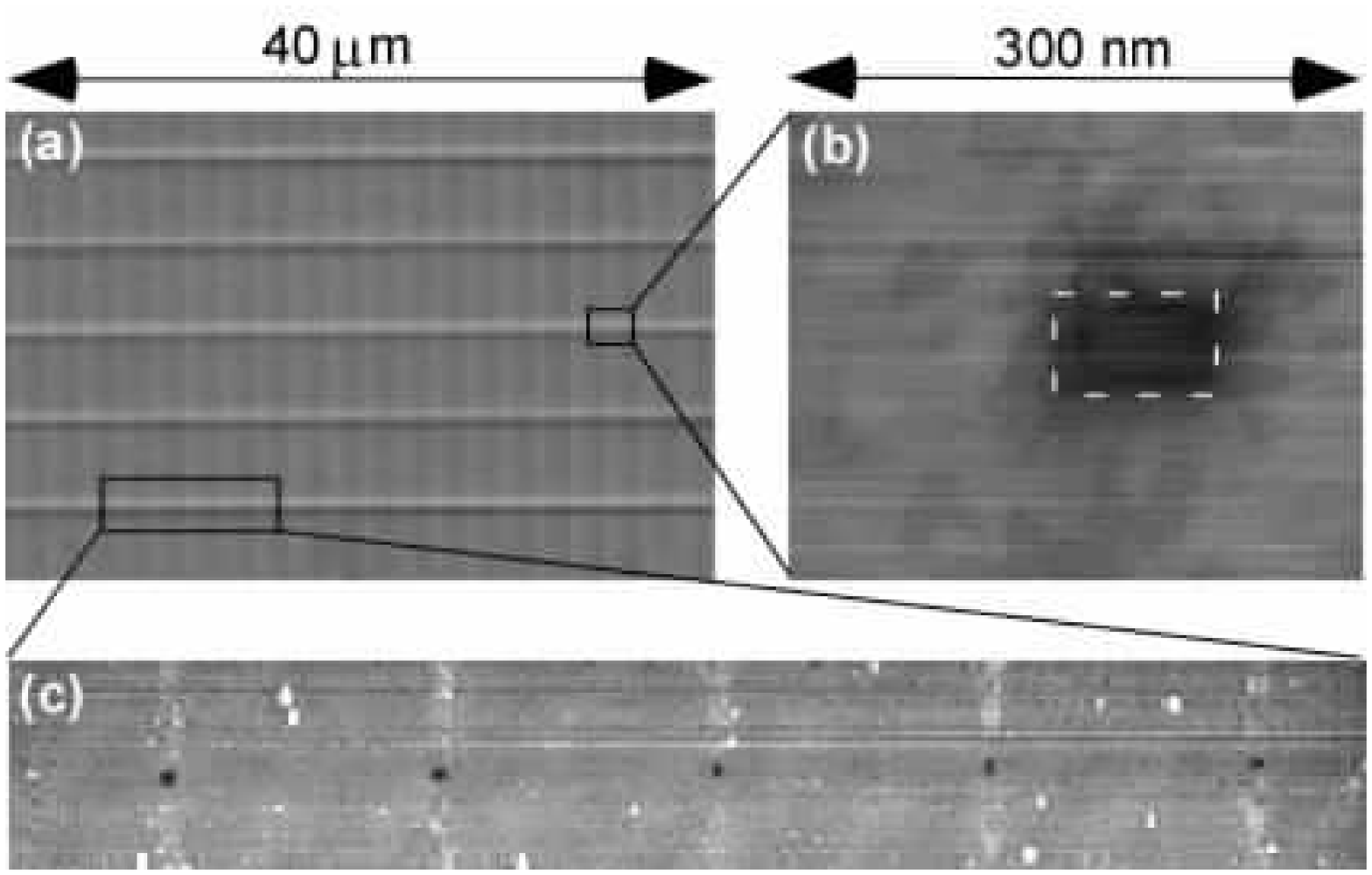}
\caption{Figure 4. (a) Trial cross-patterning results of
EBL-defined lines in a trilayer resist, as detailed in the text.
Patterning of the bottom PMGI layer is just visible as feint
vertical lines. (b) and (c) AFM images of pits etched at the
intersections of lines shown in (a).}
\end{figure}

Following ion implantation, the bilayer is fully developed and any
PMGI remaining in the cavity is removed with a solvent which does
not affect the bilayer. Fig. 3(d) shows the profile of a typical
bilayer cavity. Triple-angle Al shadow evaporation is then
performed with in-situ oxidation between the first and second
angle evaporations to form the Al/Al$_{2}$O$_{3}$ tunnel junctions
required for SET operation. Removal of any remaining resist then
leaves the structure shown in Fig. 2(a) on the substrate surface.

\begin{figure}[b]
\centering
\includegraphics[width=7.5cm]{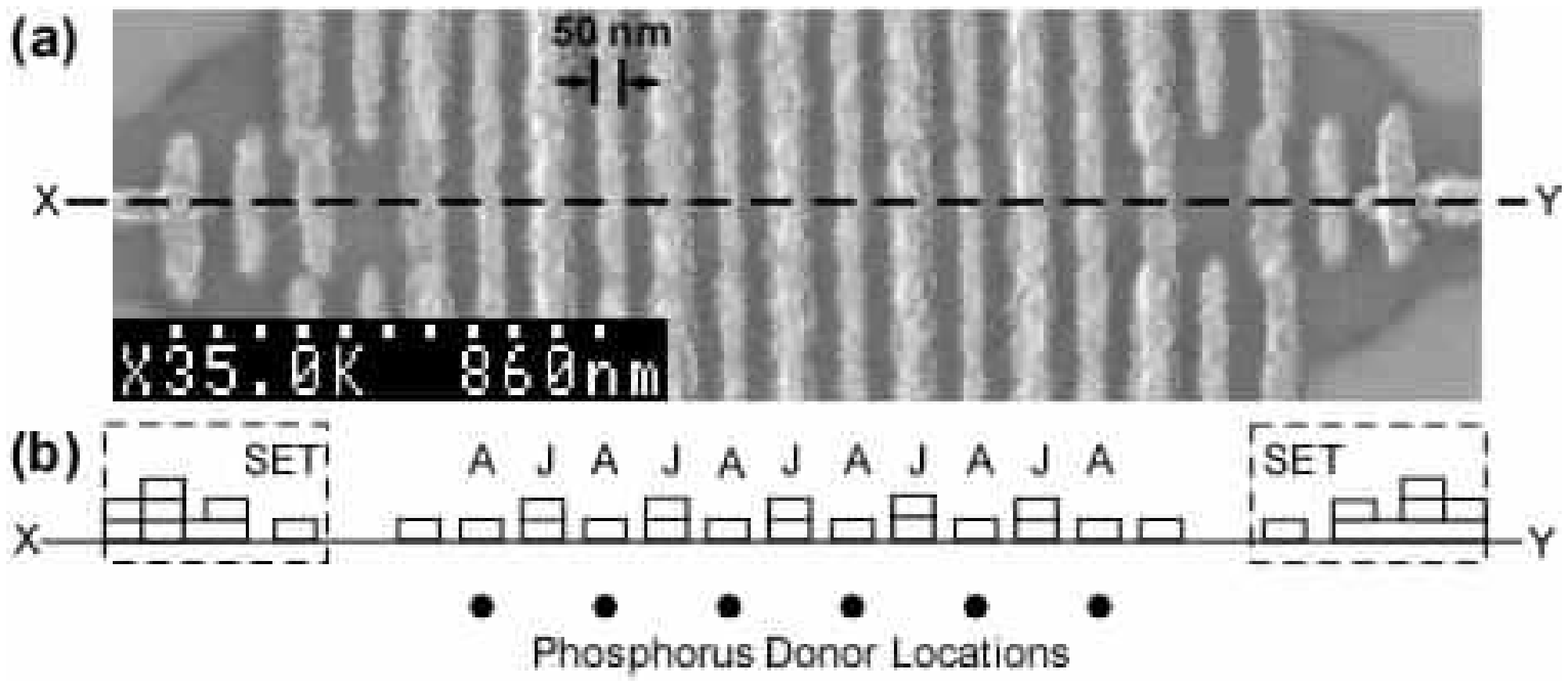}
\caption{Figure 5. (a) SEM image of gate and SET array for a
six-donor SSQC device produced by triple-angle evaporation. (b)
Cross section of the device in (a) through the line marked XY.}
\end{figure}

The triple-angle evaporation process described above has been
demonstrated using Pattern A on a bilayer resist, producing the
device shown in Figure 5(a). Fig. 5(b) shows a side-view schematic
of the same device indicating the various metal layers deposited
during the three evaporation steps. The intended donor sites are
indicated although ion implantation was not carried out on this
trial device. The 50 nm-wide lines in Fig. 5(a) are the A-gates,
which lie directly above the intended donor sites. The J-gates on
this device are slightly wider than the A-gates due to imperfect
overlaying of metal lines. The gate spacing of this device is
sufficiently small to fabricate an electron-spin SSQC \cite{a6},
however, it will be necessary to reduce metal linewidths to 5-10
nm to realize a nuclear-spin-based device \cite{a4,a5}. Such
linewidths have been realized using single layer EBL resists
\cite{a14} and work is underway to achieve this using our
multilayer process.

\begin{figure}[b]
\centering
\includegraphics[width=7.5cm]{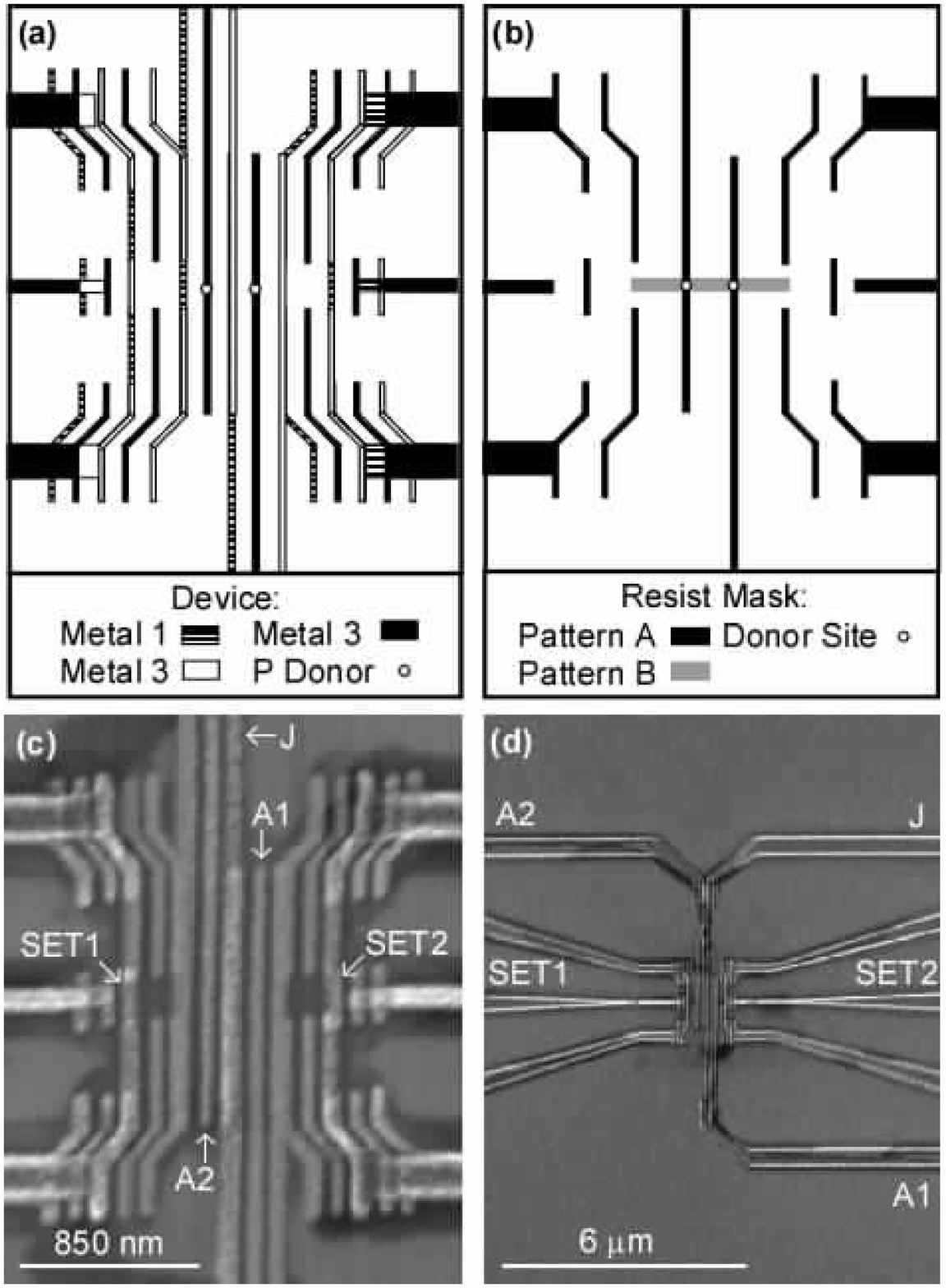}
\caption{Figure 6. (a) Schematic of a two-donor device. (b) The
two EBL patterns required to form the structure shown in (a). (c)
and (d) SEM images of the gate and SET array produced by
triple-angle evaporation.}
\end{figure}

We are currently integrating ion implantation into the process to
produce a fully configured few-qubit SSQC device. In the first
instance, a two-donor device (Fig. 6a) will be implanted. Fig.
6(b) shows the EBL patterns required to produce such a device.
Fig. 6(c) is an SEM image of a demonstration metallisation of the
gate and SET array formed by triple-angle shadow evaporation using
Pattern A. The same device is shown on a larger scale in Fig. 6(d)
revealing connections leading out to macroscopic bond pads.

\begin{figure}[t]
\centering
\includegraphics[width=7.5cm]{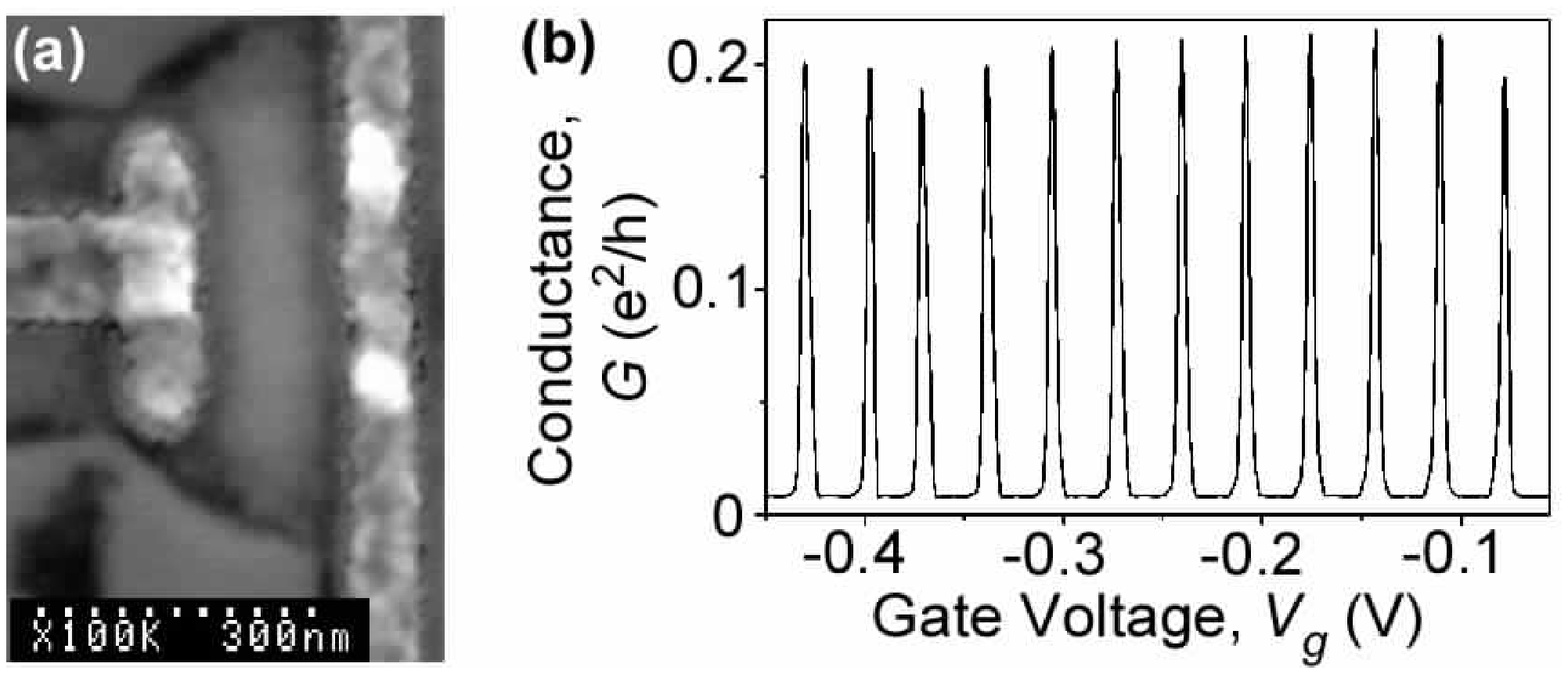}
\caption{Figure 7. (a) SEM image of an Al/Al$_{2}$O$_{3}$ SET. (b)
Coulomb blockade oscillations measured on a typical SET at 30 mK.}
\end{figure}

Figure 7(a) shows a detailed SEM image of a typical SET fabricated
by double-angle shadow evaporation on a bilayer resist. Fig. 7(b)
shows the clear Coulomb blockade oscillations measured in such
devices. The high transconductance $(dG/dV_{g})$ demonstrates the
sensitivity of these devices to single charge transfer events, and
thus their suitability for use as qubit read-out devices in the
Kane scheme. Recently, we have used two such devices in a
$twin$-SET configuration to measure controlled electron transfer
between two metallic dots, joined by a tunnel barrier \cite{a15}.

Although here we have focused on a particular process
implementation using PMGI, PMMA and PMMA/PMAA, there are a number
of possible variations. To obtain the highest possible aspect
ratios for the ion implantation channels, dry etching could be
used in place of the wet development procedures discussed above.
For example, instead of a standard bilayer structure, Resist A
could be composed of a Ge layer sandwiched between two organic
resists \cite{a16}. Pattern A could be exposed in the top resist
layer and developed, then reactive ion etching with CF$_{4}$ gas
used to remove the exposed Ge \cite{a16}. Next, another gas could
be used to anisotropically etch the underlying organic resist,
thereby defining vertical-walled ion implantation channels. In
another variation, the order of Resists A and B could be reversed.
After ion implantation, Resist B would be removed, allowing for
complete development of Resist A, followed by triple-angle shadow
evaporation.

In conclusion, we have designed a process involving two EBL
exposures, ion implantation and triple-angle shadow evaporation
capable of producing self-aligned gate, SET and donor structures
for a few-qubit SSQC. The process can utilize a variety of
different resists, development and etch procedures. We have made
preliminary test structures comprising gates and read-out SETs and
have demonstrated that cavities for ion implantation may be formed
in an organic trilayer resist structure. Work on incorporating ion
implantation into our process is currently underway.

\section*{Acknowledgement} This work was supported by the Australian
Research Council, the Australian Government and by the U.S.
National Security Agency (NSA), Advanced Research and Development
Activity (ARDA) and the Army Research Office (ARO) under contract
number DAAD19-01-1-0653.

\end{document}